# Superluminal propagation and broadband omnidirectional antireflection in optical reflectionless potentials

L. V. Thekkekara[1], Achanta Venu Gopal[1*], Sachin Kasture[1], Gajendra Mulay[1], S. Dutta Gupta[2]

[1]DCMP&MS, Tata Institute of Fundamental Research, Homi Bhabha Road, Mumbai 400 005

[2]School of Physics, University of Hyderabad, Gachibowli, Hyderabad 500 046

Reflectionless potentials (RPs) represent a class of potentials that offer total transmission in the context of one dimensional scattering. Optical realization of RPs in stratified medium can exhibit broadband omnidirectional antireflection property. In addition to the antireflection property, RPs are also expected to demonstrate negative delay. We designed refractive index profiles conforming to RPs and realize them in stratified media consisting of $Al_2O_3$ and $TiO_2$ heterolayers. In these structures we observed < 1% reflection over the broad wavelength range of 350 nm to 2500 nm for angles of incidence 0 - 50 degrees. The observed reflection and transmission response of RPs are polarization independent. A negative delay of about 31 fsec with discernible pulse narrowing was observed in passage through two RPs. These RPs can be interesting for optical instrumentation as broadband, omni-directional antireflection coatings as well as in pulse control and transmission applications like delay lines.

Reflectionless potentials (RPs) and their mathematical formalism were first discussed by Kay and Moses[1]. More recently Dutta Gupta and Agarwal have proposed a way to design optical RPs for antireflection (AR) coatings[2]. There are, in principle, infinite potentials that result in zero reflection[2,3]. Numerical simulations of Gaussian pulse transmission through RPs showed that pulse narrowing and negative group delay are possible[4]. The negative group delay can be estimated from the Wigner phase time which is the frequency derivative of the phase of the transmitted light[5]. Note that, superluminal transmission does not violate causality and has been demonstrated in several systems like atomic gases[6-9], photonic crystals[10], fiber gratings[11], corrugated waveguides[12], negative index materials[13,14], absorbing media[15,16], saturable absorbers[17-22]. There has been an experimental demonstration of a discrete coupled waveguide system conforming to reflectionless Ablowitz-Ladik soliton potential[23]. However for practical applications, the most notable application of RPs can be as an AR coating realized through a stratified medium. A coarse grained realization of a RP offered a reflectivity of < 1% over 375 nm to 800 nm wavelength range[24].

AR coatings are, in general, interference based which work on the principle of cancellation of reflections by destructive interference over narrow regions of wavelength and incidence angles or graded index which work over larger ranges of wavelength and angles due to gradual change in the refractive index[25-35]. Schallenberg's article gives a comprehensive list of references on AR coatings[24]. More recently, using bio-inspired guanine crystals, surface Mie resonators and localized plasmon resonances in metal nanoparticles, broadband omnidirectional AR coatings have been reported[27,30,36]. We design a RP and experimentally demonstrate it by making a refractive index profile conforming to the designed RP. In this, we demonstrate < 1% transmission over the broad wavelength range and negative delay when a short pulse transmits through these potentials. These have applications in optical communication, optical instrumentation and light harvesting.

The propagation equation for electric field when compared with the Schrodinger equation gives the relation between the potential and the refractive index profile as, $V(z) = k_o^2 \varepsilon_s - k_o^2 \varepsilon(z)$ where $k_o = \omega/c$, $\varepsilon_s$ is the dielectric constant far away from the potential profile given by $\varepsilon(z)$. $V(z)$ is reflectionless if any wave with positive energy passes through the potential. It has been shown that such a reflectionless potential for TE polarized light is almost reflectionless for TM as well[2]. In addition, the reflection at shorter wavelengths is shown to be not too different from the longer wavelengths thus resulting in broad wavelength range anti-reflection coatings. Design flow for such anti-reflection coating starts by considering a set of linear equations given by, $\sum M_{ij} F_j(z) = -A_i e^{\kappa_i z}$ where summation runs over j=1,N with $A_i$ and $\kappa_i$ being positive arbitrary constants for I = 1, N. The reflectionless potential V(z) is given in terms of the determinant (D) of the matrix $M_{ij}$ as[2],

$$V(z) = -2 \frac{d^2}{dz^2}[\log(D)] \quad (1)$$

From this the refractive index profile is given by,

$$n^2(z) = n_s^2 + \frac{2}{k_0^2} \frac{d^2}{dz^2}[\log(D)] \quad (2)$$

The value of D depends on the matrix $M_{ij}$ and thus on the choice of parameters $A_i$ and $\kappa_i$. For 1 parameter family with $A_1 = 2\kappa_1$, V(z) conforms to the Poschl-Teller potential. It may be seen that with the N (dimension of the parameter set) increasing we can get more and more complex profiles that can result in infinitely many reflectionless potential profiles. The choice of the parameters thus can be chosen so that the reflection potential can be achieved with the available materials.

We use 4-parameter family in designing the reflectionless potential. Solid line in Figure 1 shows the calculated profile for $[A_1,A_2,A_3,A_4]=$ [5.8,0.7,1.0,0.1] and $[\kappa_1,\kappa_2,\kappa_3,\kappa_4]$=[2.9,0.6,0.6,0.1]. In addition, we consider the role of substrate and finite potential profile thickness. Considering that we have $A_1 = 2\kappa_1$ and the values of $n_{max}$ (2.25) and $n_{min}$ (1.67) from the potential profile, we get the design wavelength to be 2.3µm. The parameters and thus the calculated refractive index profile are chosen such that it can be realized by a heterolayer consisting of pairs of $Al_2O_3$ and $TiO_2$ layers. The designed profile was realized with 43 periods of 24 nm thick each. By varying the relative thickness of constituents from pair to pair we achieve the required refractive index profile. Circles in Figure 1 show the average refractive index of each period that is spatially spread over 12 nm on each side of the mean position plotted. The size of one period and the number of periods are chosen such that the peak height and the full width at half maximum of the calculated and realized patterns match. To reduce inhomogeneity and to attain symmetric structure, we deposited half the pattern on optically flat quartz pieces. Two halves put together form the full reflectionless potential with quartz on either side.

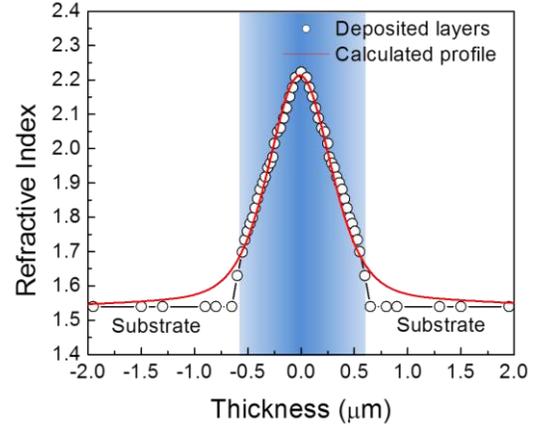

Figure 1 Reflectionless refractive index profile that is calculated (red line) and the deposited heterolayer (circles) between two quartz substrates are shown. Blue shaded region shows the schematic of the potential profile.

Deposition of heterolayer was done by electron beam evaporation on quartz substrates or rf-magnetron sputtering on ITO coated glass. $Al_2O_3$ and $TiO_2$ individual layers were first characterized by profilometer and AFM for the deposition rate as well as surface quality. This was followed by refractive index measurement by spectroscopic ellipsometer and optical transmission and reflection spectral measurements. The deposition rates were 8.4 nm/min for $Al_2O_3$ and 4.2 nm/min for $TiO_2$.

Surface roughness of the layers is about 1 nm for both. Refractive indices are 1.63 for $Al_2O_3$ (136 nm thick) and 2.33 for $TiO_2$ (76 nm thick) at 635 nm.

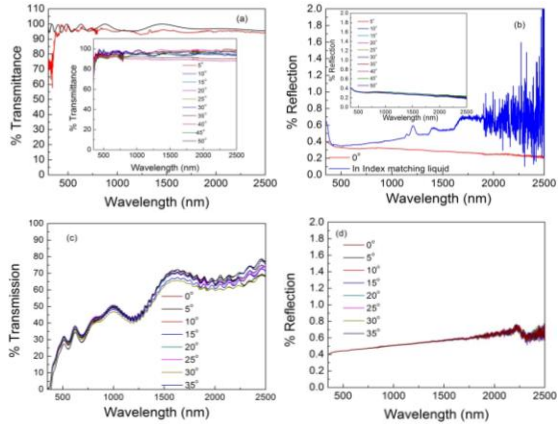

Figure 2 (a) Transmission spectra of RP is shown measured (red line) and calculated (black line). Inset shows the angle dependence of the % transmission as a function of wavelength. (b) Reflection spectra at normal incidence in air (red line) corrected for air-quartz interface reflection and that measured in index matching liquid (blue line). Inset shows the angle dependence of the normalized reflection spectra. Angle dependence of (c) % transmission and (d) % reflection a RP deposited on patterned substrate.

Transmission and reflection efficiency through one full pattern were measured by splitting the incident beam into two parts. One beam falls on the sample and the other, reference beam, on quartz pieces without the heterolayer to take into account the reflection at the air-quartz interface. T and R as a function of wavelength for normal incidence are plotted in Figure 2. Black line shows the calculated T and red line shows the measured data. Excellent match of the Fabry-Perot oscillations in the calculated spectrum for the exact heterolayer structure and the measured data shows that the heterolayers conform to the potential designed. In addition, the reflectance data shows the broadband anti-reflection nature of the heterolayer structure by showing less than 1% reflection in the 350 nm – 2500 nm wavelength range. Angle dependence up to 50° incidence angle (limited by the measurement geometry) is plotted in Figure 2. We also performed measurements in index matching liquid (Cargille Labs, USA) which showed < 1% transmission over the broad wavelength region. We prepared two more structures with different full width at half maxima by having base periods of 18 nm and 14 nm, respectively. Both these structures also show very low reflectivity over the broad wavelength region (data not shown).

To overcome the reflection at the air – quartz and air – $Al_2O_3$ interfaces of the above two structures, we prepared patterned quartz substrates on which RP was deposited by e-beam evaporation. A 1mm x 1mm area on the quartz substrate was patterned with a 2-dimensional aperiodic array of air holes with conical depth profiles. By electron beam lithography the pattern was written in gold mask which was then transferred to the quartz substrate by reactive ion etching in $CHF_3/O_2$ plasma. About 140 nm deep holes of $80 \pm 10$ nm diameter are achieved with a taper. Air hole density is $1.4x10^6$ /$mm^2$. Measured % transmission and reflection on RP made on patterned substrate are shown in Figs. 2c and 2d, respectively. Transmission seems to be affected by the patterned substrate but the reflection is very low over the broad wavelength range.

Another pertinent feature of reflectionless potentials is the negative group delay of a pulse passing through RP, which were reported by Kiriuscheva and Kuzmin using numerical simulations[3]. In the stationary phase approximation, the delay can be estimated by the frequency derivative of the phase of the transmitted light also known as the Wigner phase time[5]. We report the computational and experimental results on negative delay in pulse transmitted through reflectionless potentials. We performed two dimensional finite difference time domain simulations (Lumerical) to study the transmission of a TM polarized, 1.8 μm beam waist, 100 fsec Gaussian pulse centered at 800nm wavelength through the structures fabricated. We used perfectly matched layer (PML) boundary conditions with mesh size of 1 nm in x and y directions. We considered $Al_2O_3$ and $TiO_2$ dispersion and considered actual layer thicknesses. Simulation results are shown in Fig. 3a for transmission through single RP. A negative delay of -0.01 psec is seen for single RP of length 1.1 μm. Experimental short pulse transmission studies were performed by time resolved transmission using fsec pulses centered at three different wavelengths of 745 nm, 800 nm and 850 nm from a Ti : Sapphire laser.

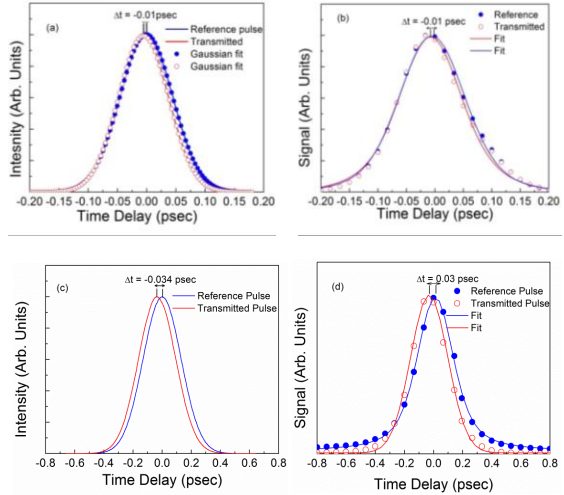

Figure 3 Reference and transmitted 100 fsec pulse profiles simulated by FDTD (a) and measured (b) for reflectionless profile between quartz substrates. Central wavelength of incident pulse is at 745 nm. Simulated (c) and experimental (d) negative delay and pulse profile for the 300 fsec pulse at 800 nm wavelength transmitted through two RPs deposited on ITO coated glass. All transmitted pulse signals are multiplied by a suitable constant to compare with the reference pulse.

Experimental result for pulse centered at 745 nm wavelength are shown in Fig. 3b which match well with the simulation result showing -0.01 psec time delay between the reference pulse and the pulse transmitted through RP deposited between two quartz pieces. Similar negative delays observed for pulses centered at 800 nm and 850nm wavelengths shows the broadband nature of the RPs. Figure 3c shows the negative delay for short pulse centered at 800 nm wavelength when transmitted through two successive RPs of total thickness 2.2 μm. With respect to the reference pulse a negative delay of 0.031 ± 0.01 psec corresponding to group index $n_g$ of -3.91 ± 1.5 was observed. In addition to the negative delay, pulse narrowing is also seen in the results.

In conclusion, reflectionless potentials have interesting properties like broadband omni-directional antireflection coatings, distortionless pulse transmission and negative delay. We demonstrate reflectionless potentials in optical domain, realized in stratified media, that exhibit superluminal transmission. 31 fsec negative delay was observed when short pulse is transmitted through two RPs. Broadband (350 nm to 2500 nm) omni-directional anti-reflection (< 1%) feature observed. These have interesting applications in optical communication, optical instrumentation as well as in light harvesting.